\documentclass[aps,pra,groupedaddress,onecolumn,nopacs,11pt,tightenlines]{revtex4-1}
\usepackage{color}
\usepackage{dcolumn}
\usepackage{graphicx}
\usepackage{amsmath}
\usepackage{amssymb}
\usepackage{epstopdf}
\usepackage{braket}
\usepackage{soul}

\newcommand{\average}[1]{\langle #1 \rangle}

\begin{document}


\title{Dynamical properties of dissipative XYZ Heisenberg lattices}


\author{R. Rota$^*$, F. Minganti$^*$, A. Biella, C. Ciuti}
\affiliation{Laboratoire Mat\'eriaux et Ph\'enom\`enes Quantiques, Universit\'e Paris Diderot, CNRS UMR 7162, Sorbonne Paris Cit\'e, 10 rue Alice Domon et Leonie Duquet 75013 Paris, France}



\date{\today}

\begin{abstract}

We study dynamical properties of dissipative XYZ Heisenberg lattices where anisotropic spin-spin coupling competes
with local incoherent spin flip processes. In particular, we explore a region of the parameter space where dissipative magnetic phase transitions for the steady state have been recently predicted by mean-field theories and exact numerical methods. We investigate the asymptotic decay rate towards the steady state both in 1D (up to the thermodynamical limit) and in finite-size 2D lattices, showing that critical dynamics does not occur in 1D, but it can emerge in 2D. We also analyze the behavior of individual homodyne quantum trajectories, which well reveal the nature of the transition. \end{abstract}


\maketitle



\section{Introduction}\label{sec:intro}

Quantum many-body physics with light has proved to be an extremely rich and interesting field of study, as it combines the complexity of condensed matter with the intrinsically out-of-equilibrium behavior of optical systems \cite{CiutiRMP13, HartmannJOpt16, LeHurCRP16, NohRPP17}.
Collective phenomena among photons, such as Bose-Einstein condensation \cite{BaumbergPRB00,StevensonPRL00,BaasPRL06,KasprzakNature06} or superfluidity \cite{AmoNatPhys09,NardinNatPhys11,SanvittoNatPhot11,AmoScience11,GrossoPRL11}, have been observed in planar semiconductor microcavities in the strong light-matter coupling regime. In these systems, the optical confinement and the nonlinearity of the media give rise to a weak photon-photon interaction, which allows the many-photon system to behave as a quantum fluid.

The appearance of strongly correlated states of light is even more evident in regimes where the interaction among photons becomes large. When the nonlinearity of the optical cavity is much larger than its dissipation rate, the presence of a single photon inside the cavity is able to effectively block the entrance of a second one. This effect, known as \textit{photon-blockade} \cite{CarmichaelPRL85,ImamogluPRL97}, has been observed experimentally at first with optical photons using a single atom in a cavity \cite{BirnbaumNat05} and is particularly strong in circuit quantum electrodynamics systems in the microwave domain \cite{LangPRL11}. Non-trivial phases can also arise when several cavities are coupled together and form a lattice of resonators \cite{You11}. For instance,  correlations can lead to a transition from a photonic Mott insulator to a superfluid \cite{GreentreeNatPhys06,HartmannNatPhys06,AngelakisPRA07,HartmannLPR08,LebreuillyPRA17}, similar to that observed with ultracold atoms confined in optical lattices \cite{GreinerNature2002,BlochRMP08}.
Interestingly, a system of coupled resonators in the photon-blockade regime arranged according a lattice geometry can be mapped into an effective spin model \cite{AngelakisPRA07,HartmannPRL07,KayEPL08}. This class of systems can be realized nowadays using different experimental platforms, such as superconducting quantum simulators \cite{HouckNP2012} or Rydberg atoms \cite{LeePRA11,QianPRA12,ViteauPRL12,QianPRA15}.

Among the collective phenomena appearing in coupled photonic lattices, dissipative phase transitions are nowadays deserving more and more attention.  Dissipative processes are usually at odds with the unitary Hamiltonian evolution of the quantum system and the competition between the incoherent and the coherent dynamics can give rise to criticality for the steady state in the thermodynamic limit \cite{KesslerPRA12}. Dissipative phase transitions have been discussed theoretically for single cavity photonic systems \cite{CarmichaelPRX15,MendozaPRA16,BartoloPRA16}, as well as for lattices of cavities with mean field methods \cite{BiondiPRA17,BiellaPRA17,SavonaPRA17} or full-size lattice simulations \cite{Foss-FeigPRA17,VicentiniarXiv17}.
An experimental observation of these critical phenomena seems feasible with state-of-the-art techniques, and some remarkable results have already been obtained \cite{FinkPRX2017,FinkarXiv17,RodriguezPRL2017}. 

In this context, the dissipative XYZ Heisenberg model  \cite{LeePRL13} has attracted a considerable attention. It describes a lattice of spins interacting via an anisotropic Heisenberg Hamiltonian coupled to an environment which forces spins to align along the $z$-axis. The single-site Gutzwiller mean-field theory predicts a rich phase diagram for the magnetic properties of the steady state of this model \cite{LeePRL13}. More refined calculations \cite{BiellaPRX16,RotaPRB17,OrusNatComm17,CasteelsarXiv17,BiellaArXiv17}, based on numerical methods including many-body correlations, have confirmed the emergence of a critical behavior in two-dimensional lattices, while the phase transition disappear when the spins are arranged according to a one-dimensional geometry. All these works, however, focussed on the calculation of steady-state properties and a full description of the dynamics of the system is still lacking. 

In this work, we explore the dynamical properties of the dissipative XYZ model in the region where a second-order phase transition from a paramagnetic to a ferromagnetic steady state has been predicted. For finite-size 1D arrays and 2D lattices, we have performed an exact integration of the master equation using the whole Hilbert space via the Wave Function Monte Carlo method \cite{MolmerJOSAB93}. Moreover, for 1D arrays of infinite length we have applied the infinite Matrix Product Operator (iMPO) technique \cite{VidalPRL07,OrusPRB08}.  

This article is organized as follows.
In Sec.~\ref{sec:math} we discuss the theoretical framework and describe the methods used for the calculations. In Sec.~\ref{sec:results} we show the main results of the work. In Sec~\ref{sec:conclusions} we draw our conclusions and present some perspectives.

\section{Mathematical Framework}\label{sec:math}

The dissipative XYZ model describes a lattice of spins interacting via an anisotropic Heisenberg Hamiltonian ($\hbar = 1$):
\begin{equation}\label{Eq:Hamiltonian}
\hat{H}=\sum_{\langle i , j \rangle} \left(J_x \hat{\sigma}^x_i \hat{\sigma}^x_j + J_y \hat{\sigma}^y_i \hat{\sigma}^y_i + J_z \hat{\sigma}^z_i \hat{\sigma}^z_j \right),
\end{equation}
where $\hat{\sigma}^\alpha_i$ ($\alpha=x,y,z$) represent the Pauli matrices acting on the $i$-th site. The sum runs over the nearest neighbour sites $\langle i , j \rangle$. The dissipative part describes incoherent spin-flip processes which tend to align a single spin towards the negative direction of the $z$-axis with a rate $\gamma$. The density matrix $\hat{\rho}(t)$ dynamics is obtained from the Lindblad master equation
\begin{equation}\label{Eq:Lindblad}
\frac{\partial \hat{\rho}}{\partial t} = \mathcal{L} [\hat{\rho}]= - i \left[\hat{H}, \hat{\rho}\right] +  \gamma \sum_j  \left( \hat{\sigma_j}^- \hat{\rho} \hat{\sigma_j}^+ - \frac{1}{2} \left( \hat{\sigma_j}^+\hat{\sigma_j}^- \, \hat{\rho} + \hat{\rho} \,  \hat{\sigma_j}^+ \hat{\sigma_j}^- \right) \right) \ ,
\end{equation}
where $\hat{\sigma}^\pm_j = (\hat{\sigma}^x_j \pm i \hat{\sigma}^y_j)/2$ are the spin raising and lowering operators acting on the $j$-th spin and $\mathcal{L}$ is the Liouvillian superoperator. The latter is non-Hermitian and has a spectrum of complex eigenvalues, defined by the equation $\mathcal{L} [\hat{\rho}_r] = \lambda_r \hat{\rho}_r$.

The dissipative XYZ model evolves towards a steady state $\hat{\rho}_{ss}$, which depends on the parameters in \eqref{Eq:Lindblad} and corresponds to the zero eigenvalue of $\mathcal{L}$ ($\partial_t \hat{\rho}_{ss} = \mathcal{L} [\hat{\rho}_{ss}]=0$). All the other eigenvalues $\lambda_r$ are such that their real part is negative and describe the relaxation dynamics of $\hat{\rho}(t)$ towards the steady state. Since a dissipative phase transition is expected to be characterized by a critical slowing down in the dynamics of the system, a particular relevance has to be given to the so-called Liouvillian gap $\lambda = \mathrm{min}_r |Re(\lambda_r)|$, which is also called asymptotic decay rate \cite{KesslerPRA12}. The emergence of a critical behavior is associated to a closing of the Liouvillian gap in the thermodynamic limit \cite{KesslerPRA12,CasteelsPRA17-2,VicentiniarXiv17}.

\begin{figure}
	\includegraphics[width = 0.9\textwidth]{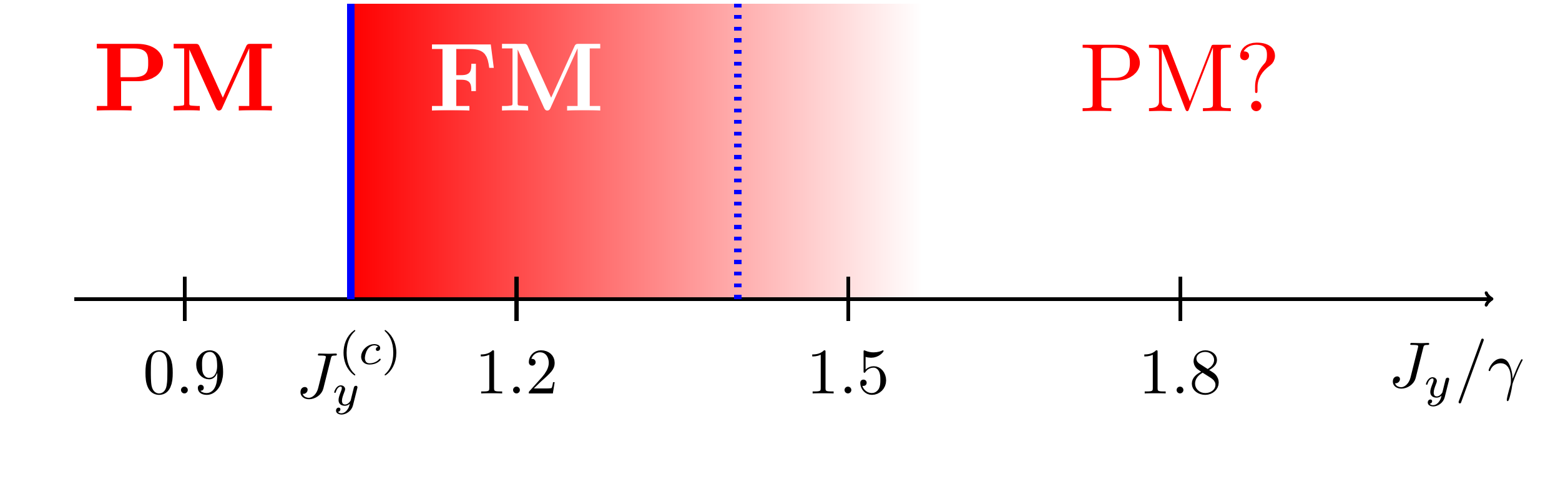}
	\caption{Phase diagram of the 2D dissipative XYZ model as a function of the normalized coupling parameter $J_y/\gamma$, with fixed $J_x/\gamma = 0.9$ and $J_z/\gamma = 1$. For $J_y \simeq J_x$, the system presents a paramagnetic (PM) steady state. At the critical value $J_y^{(c)}$, the system undergoes a phase transition towards a ferromagnetic (FM) steady state. Different estimations for this critical value are: $J_y^{(c)}/\gamma = 1.039$ from Ref. \cite{LeePRL13}, $J_y^{(c)}/\gamma = 1.04 \pm 0.01$ from Ref. \cite{BiellaPRX16}, $J_y^{(c)}/\gamma = 1.07 \pm 0.02$ from Ref. \cite{RotaPRB17} and $J_y^{(c)}/\gamma = 1.0665 \pm 0.0005$ from Ref. \cite{BiellaArXiv17}. At larger values of $J_y$, the nature of the steady state is still under debate: Ref. \cite{BiellaPRX16} predicts the existence of a second critical point $J_y^{(c,2)}/\gamma = 1.40$ (dashed blue line in the figure), above which the steady state is paramagnetic, but Ref. \cite{RotaPRB17} does not show any evidence of a phase transition close to this value.}
	\label{fig:PhaseDiagram}
\end{figure}

The Lindblad master equation (Eq. \ref{Eq:Lindblad}) is invariant under a $\pi$-rotation of all the spins around the $z$-axis ($\hat{\sigma}_i^{x} \to - \hat{\sigma}_i^{x}$, $\hat{\sigma}_i^{y} \to - \hat{\sigma}_i^{y}$ $\forall i$). In the thermodynamic limit, the $\mathbb{Z}_2$ symmetry associated to this transformation may spontaneously break, resulting in the appearance of several magnetic phases for the steady state of the model. In this work, we will focus on a particular regime where previous calculations have predicted a transition from a paramagnetic phase with no magnetization in the $xy$ plane ($\average{\hat{\sigma}_x} = \textrm{Tr} (\hat{\rho}_{ss} \hat{\sigma}^x_j) = 0$ , $\average{\hat{\sigma}_y} = \textrm{Tr} (\hat{\rho}_{ss} \hat{\sigma}^y_j) = 0$) to a ferromagnetic phase with finite magnetization in the $xy$ plane ($\average{\hat{\sigma}_x} \ne 0$ , $\average{\hat{\sigma}_y} \ne 0$) \cite{LeePRL13,BiellaPRX16,RotaPRB17,OrusNatComm17,CasteelsarXiv17,BiellaArXiv17} (see Fig. \ref{fig:PhaseDiagram}).

From a computational point of view, the numerical solution of the master equation \eqref{Eq:Lindblad} is a formidable task when considering extended lattices. The corner-space renormalization method \cite{FinazziPRL15}, which has shown the criticality of several steady-state observables in 2D lattices \cite{RotaPRB17},  does not give access to the dynamic properties of the system. For small systems with a number $N < 10$ spins, the problem can be solved via a standard Runge-Kutta integration of Eq. \eqref{Eq:Lindblad}. For $10 \leq N \leq 16$, instead, we have solved the master equation stochastically via the Wave Function Monte Carlo method \cite{MolmerJOSAB93}. This method describes the time evolution of the open quantum system in terms of a set of $N_T$ pure states $| \Psi_k (t) \rangle$ (usually called {\it quantum trajectories}), obtained independently according to a stochastic evolution protocol \cite{HarocheBOOK,CarmichaelBOOK,WisemanBOOK,DaleyAP14}. The density matrix is retrieved by averaging over the $N_T$ sampled trajectories, according to the formula $\hat{\rho}(t) = 1/N_T \sum_{k=1}^{N_T} |\Psi_k(t) \rangle \langle \Psi_k(t)|$. The computational advantage of this method is clear, as it allows to study the evolution of the open system dealing with pure states (which are vectors of size $2^N$), instead of the density matrix (which has size $2^N \times 2^N$).

It is important to notice that quantum trajectories are useful not only to reduce the complexity of the integration of the Lindblad master equation \eqref{Eq:Lindblad}, but their analysis of can also provide insightful results about the nature of the dissipative phase transition \cite{Foss-FeigPRA17,VicentiniarXiv17}. To this aim, we have investigated the stochastic evolution of individual quantum trajectories for the dissipative XYZ model, obtained according to the \textit{homodyne} protocol described by the following equation:
\begin{equation}\label{Eq:StocSchHomodyne}
\lvert{\Psi_k(t + dt)}\rangle =  \left \{ -i\, dt\, \hat{H} + \sum_{j} \sqrt{\gamma}
\left [  \hat{\sigma}_j^- - \frac{s_j(t)}{2}\right ] dW_j(t) 
- \frac{\gamma}{2} \left [ \hat{\sigma}_j^+ \hat{\sigma}_j^-
-  s_j(t) \hat{\sigma}_j^-  + \frac{s_j(t)^2}{4} \right ] dt
\right \}  |{\Psi_k(t)}\rangle,
\end{equation}
where $s_j(t) =\langle \Psi_k(t) | \hat{\sigma}_j^x |\Psi_k(t) \rangle$ and $dW_{j}$ are stochastic Wiener increments with zero expectation value, variance equal to $\sqrt{dt}$ and uncorrelated among the different spins (the detailed derivation can be found, e.g., in~\cite{WisemanBOOK}). Contrarily to the master equation \eqref{Eq:Lindblad}, the stochastic equation in \eqref{Eq:StocSchHomodyne} does not conserve the $\mathbb{Z}_2$ symmetry of the Liouvillian superoperator, due to the presence of the terms $s_j(t)$. Therefore, by studying the time evolution of the magnetic order parameter over an individual quantum trajectory, it is possible to reveal the emergence of different magnetic phases, when we change the parameters of the system. Nevertheless, the symmetry of the Liouvillian is restored when we consider the density matrix, obtained by averaging over many trajectories.

Alternative approaches for the simulation of 1D arrays are based on tensor networks techniques \cite{OrusAP14} making use of the Matrix Product Operator (MPO) ansatz for the density matrix \cite{VestraetePRL04,ZwolakPRL04} (see for example Refs. \cite{JoshiPRA13,BonnesPRA14,BiellaPRA15,BiondiPRL15,BiellaPRX16,BiellaPRA17}). The MPO ansatz for the many-body mixed state 
can be controlled by changing a single parameter, i.e. the bond-link dimension $\chi$: the more $\chi$ increases, the more non-local quantum correlations can be encoded. 
The dynamics of the open system is obtained via a time-evolving block decimation scheme \cite{VidalPRL03,VidalPRL04}. 
In the case of translational invariant systems, the MPO ansatz and the time evolution procedure can be further simplified leading to the infinite MPO (iMPO) representation \cite{VidalPRL07,OrusPRB08}, which allows to directly access the thermodynamic limit of an infinite number of sites.
Very recently, this technique has been extended to the case of 2D lattices \cite{OrusNatComm17} although with a very reduced bond dimension.

\section{Results and Discussion}\label{sec:results}

\begin{figure}[t]
\begin{center}
\includegraphics[width = 0.9\textwidth]{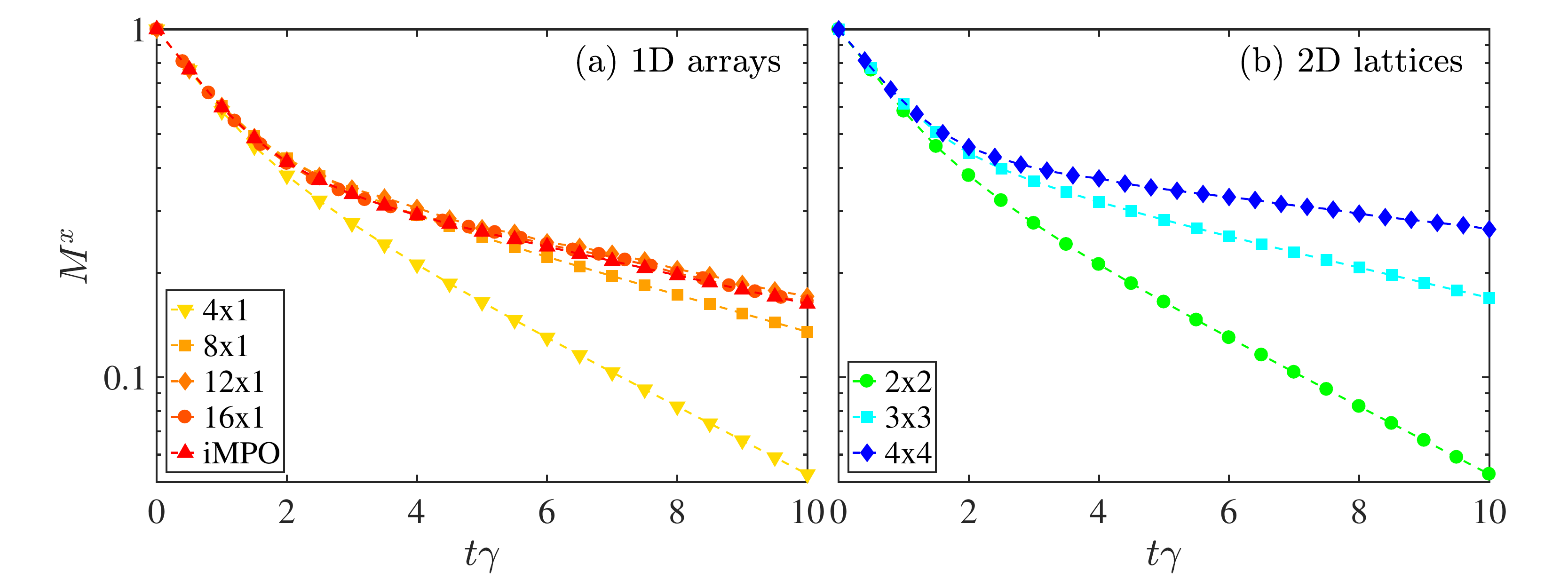}
\includegraphics[width = 0.9\textwidth]{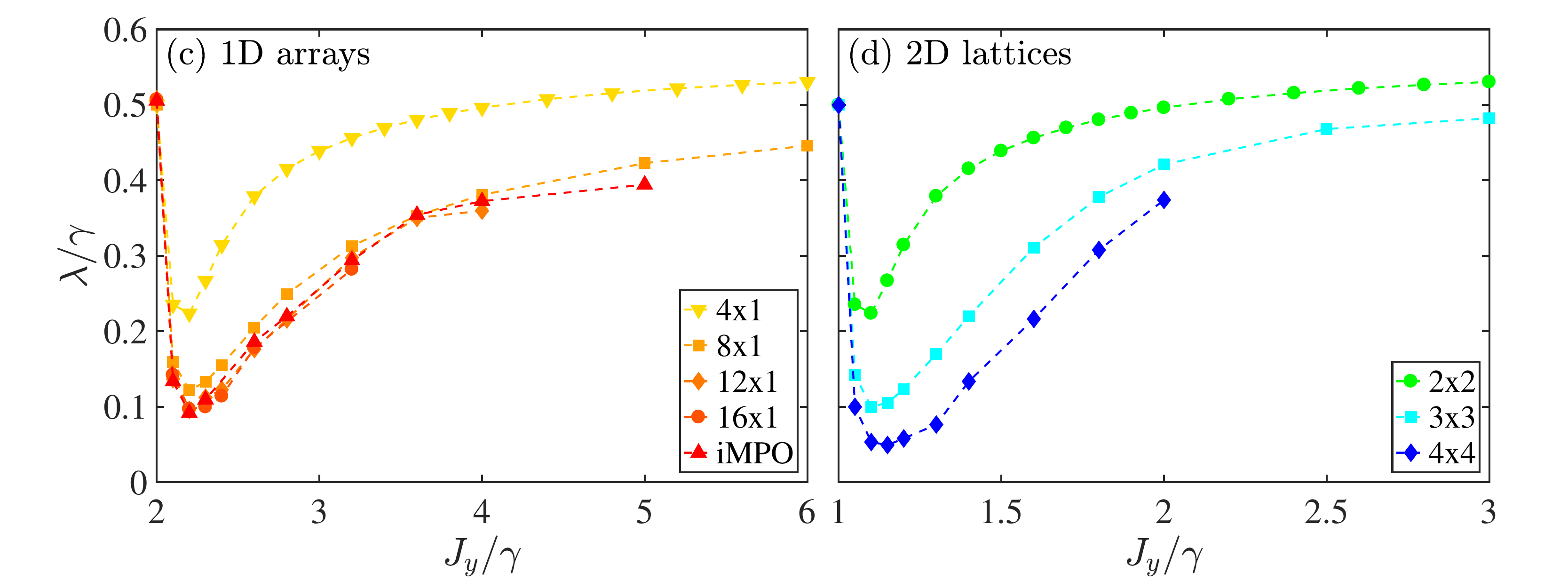}
\end{center}
\caption{Top panels: time dependence of the averaged magnetization $M^x(t)$ in 1D arrays (panel (a)) and in 2D lattice (panel (b)) of different size. Parameters: $J_x/\gamma = 1.8$, $J_y/\gamma = 2.2$ and $J_z/\gamma = 2$ for the 1D results in panel (a); 
$J_x/\gamma = 0.9$, $J_y/\gamma = 1.1$ and $J_z/\gamma = 1$ for the 2D results in panel (b). 
Lower panels: Liouvillian gap as a function of the coupling parameter $J_y$ in 1D arrays (panel (c)) and 2D lattices (panel (d)). The other parameters are: $J_x/\gamma = 1.8$ and $J_z/\gamma = 2$ for the 1D results; $J_x/\gamma = 0.9$ and $J_z/\gamma = 1$ for the 2D results.}\label{fig:LiouvGap}
\end{figure}

We start our discussion on the dynamics of the dissipative XYZ model by studying the time evolution of the average lattice magnetization $M^{x}(t) = \sum_i {\rm Tr} \left[\hat{\rho}(t) \hat{\sigma}^x_i\right]/N$, $N$ being the number of spins in the lattice. In Fig. \ref{fig:LiouvGap}, we plot $M^{x}(t)$ for a fixed choice of the parameters of the Hamiltonian \eqref{Eq:Hamiltonian} in vicinity of the critical point, for spin systems of different size, both with 1D (Fig. \ref{fig:LiouvGap}-(a)) and 2D geometry (Fig. \ref{fig:LiouvGap}-(b)).
In all these calculations, the master equation has been solved assuming an initial configuration where all the spins point along the positive direction of the $x$-axis (therefore $M^{x}(t=0) = 1$) and imposing periodic boundary conditions to the finite-size lattice.

For $t \gtrsim 5 \gamma$, all the curves $M^{x}(t)$ decay exponentially towards the steady-state expectation value $M^{x}_{ss} = 0$ (notice that we have $M^{x}_{ss} = 0$ for all the values of the parameters since we do not break explicitly the $\mathbb{Z}_2$ symmetry of the Liouvillian superoperator in our simulations). The presence of an asymptotic exponential behavior for $M^{x}(t)$ indicates that, at large times, the dynamics of the system can be described uniquely in terms of the eigenstate associated to the Liouvillian gap. The density matrix can be approximated as $\hat{\rho}(t) = \hat{\rho}_{ss} + A \hat{\rho}_1 e^{-\lambda t}$, where $A$ is a real number depending on the choice of the initial configuration. From our results, we notice also that the dynamics gets slower when increasing the size of the system, both in 1D arrays and in 2D lattices (respectively Fig. \ref{fig:LiouvGap}(a) and Fig. \ref{fig:LiouvGap}-(b)). In 1D arrays the decay rate saturates when the size of the system increases. For an array with $16$ sites the decay curve is nearly indistinguishable from what obtained for an array of infinite length (obtained via the iMPO technique). Instead, in 2D lattices no saturation of the decay rate is observed. 

By fitting the curves for $M^x(t)$ at large $t$ with a simple exponential, we can extract the value of the Liouvillian gap $\lambda$. The results for $\lambda$ obtained with this procedure have been successfully benchmarked against those calculated with an exact diagonalization of the Liouvillian superoperator in small systems ($4 \times 1$ array and $2 \times 2$ lattice). In Fig. \ref{fig:LiouvGap}-(c,d) we plot $\lambda$ as a function of the normalized coupling parameter $J_y/\gamma$ (the other coupling parameters $J_x/\gamma$ and $J_z/\gamma$ are kept fixed). Both in the 1D and in the 2D case, all the curves $\lambda(J_y)$ present a minimum close to the critical value of $J_c$, indicating a slowing down in the dynamics of the system. Nevertheless, we clearly notice that this slowing down is not critical in 1D systems. Indeed, the results for $\lambda(J_y)$ in the largest 1D systems (with $N \ge 12$) overlap and are in good agreement with the prediction for the infinite array obtained with iMPO \footnote{The accuracy of the iMPO data is checked by increasing the bond-dimension $\chi$ until the convergence is reached (in our calculation, convergence is obtained with $\chi = 80$)}, showing a finite value of the Liouvillian gap.  Instead, in 2D systems, the minimum of $\lambda(J_y)$ becomes smaller and smaller when the size of the lattice increases. This behavior is consistent with a closure of the Liouvillian gap in the thermodynamic limit. 

In order to better characterize the behavior of the 2D system across the critical point, we study the average magnetization of the lattice $M^x_\Psi (t) = \langle \Psi(t) | \sum_i \sigma^x_i |\Psi(t) \rangle/N$ along a single trajectory $\ket{\Psi(t)}$. To this extent, we have computed $\ket{\Psi(t)}$ following the homodyne protocol in Eq. \eqref{Eq:StocSchHomodyne} in 2D lattices of different sizes, for several values of the parameter $J_y$, starting from an initial configuration where all the spins are aligned along the $z$-axis. Convergence of the time integration of Eq. \eqref{Eq:StocSchHomodyne} has been carefully checked, requiring a time step  $dt \simeq (1000 \gamma)^{-1}$.

\begin{figure}[t]
\begin{center}
\includegraphics[width = 0.7\textwidth]{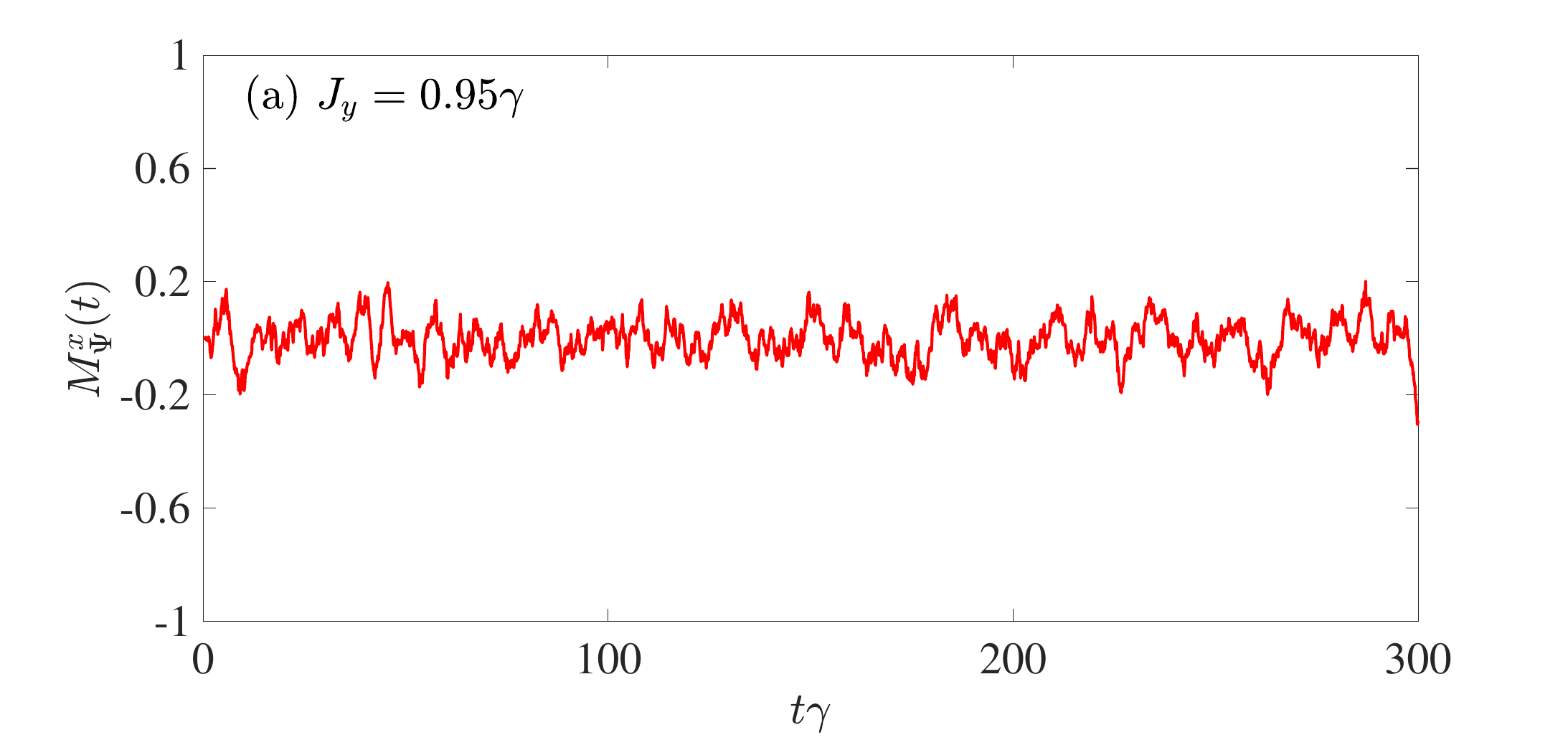}
\includegraphics[width = 0.7\textwidth]{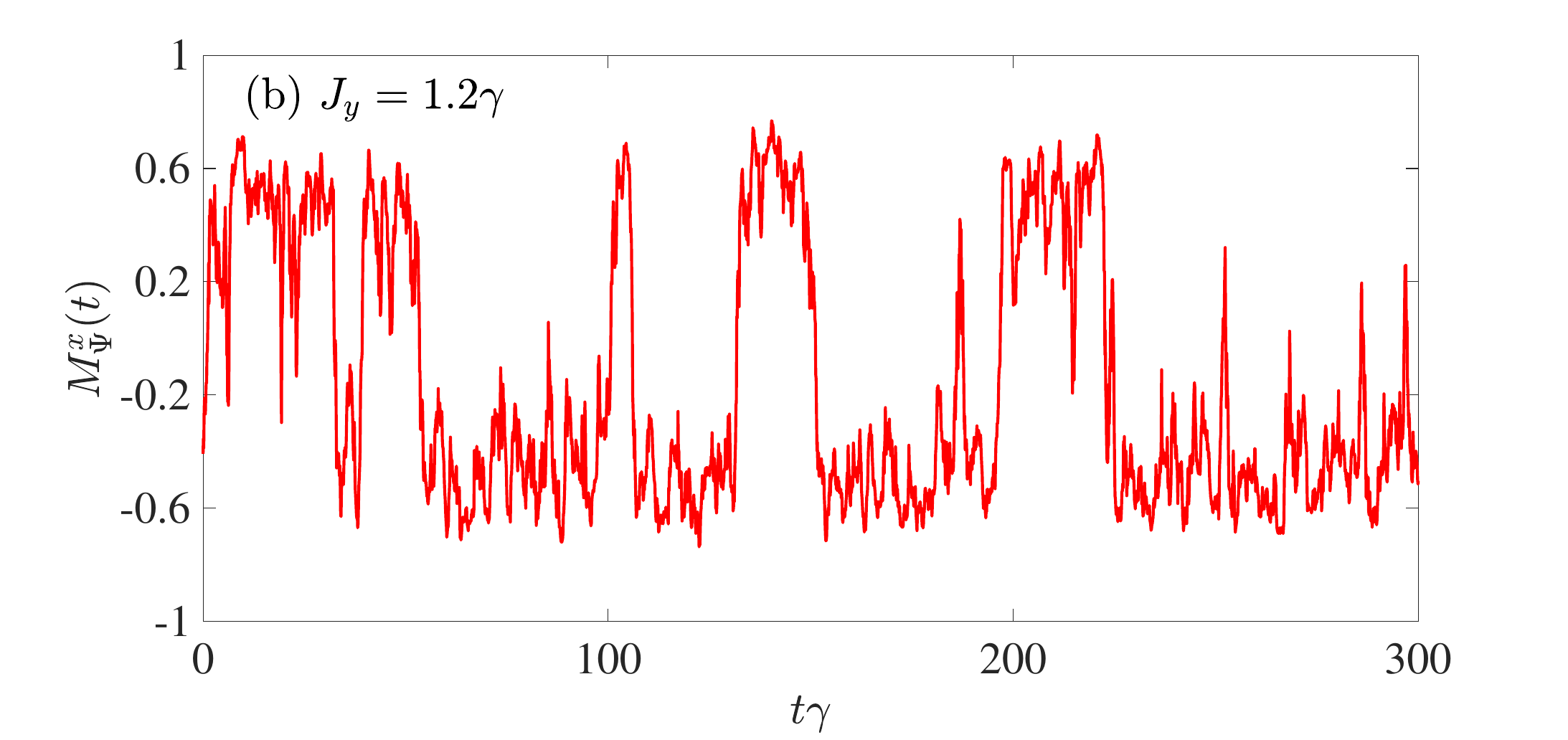}
\includegraphics[width = 0.7\textwidth]{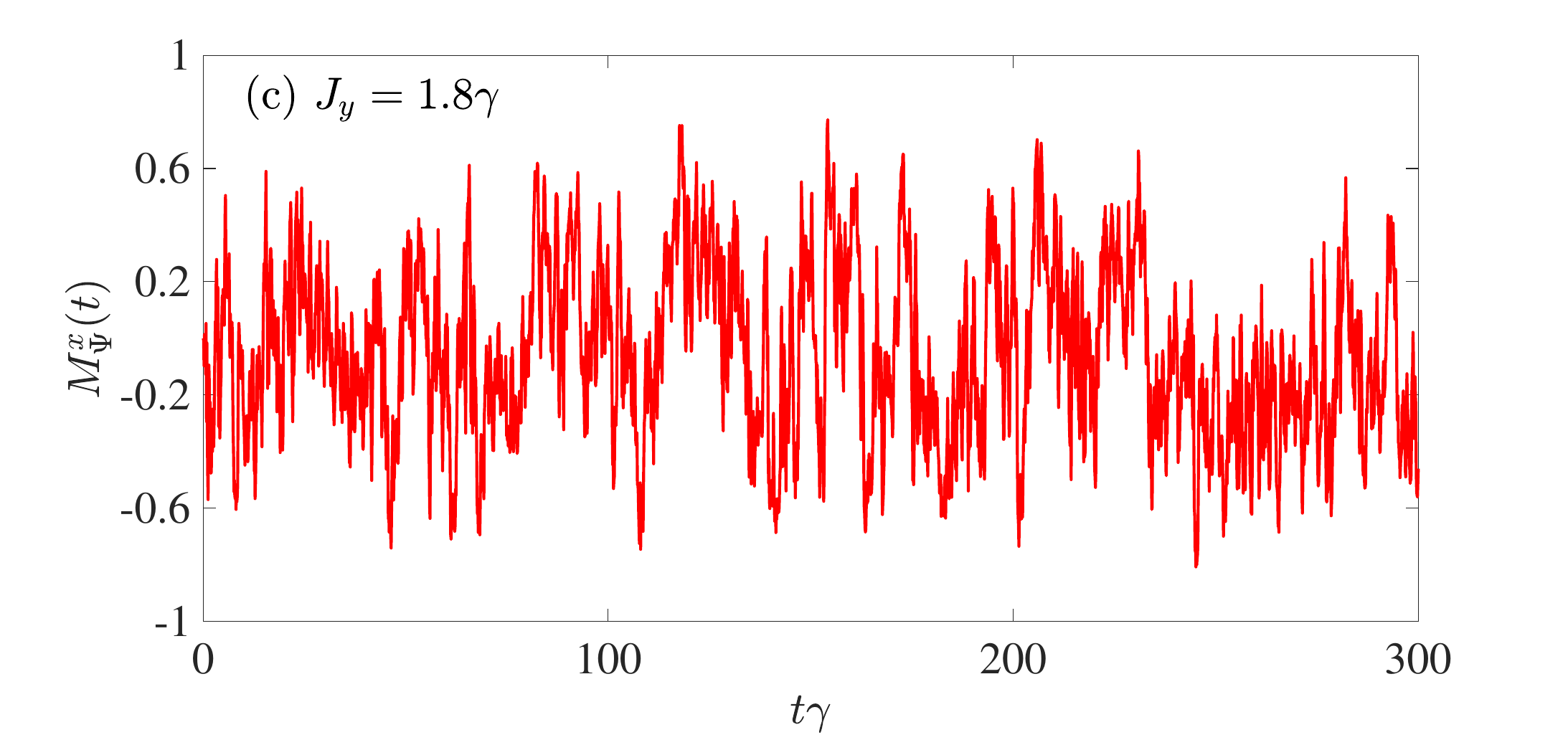}
\end{center}
\caption{Average magnetization $M^x_\Psi$ calculated for a single homodyne quantum trajectory as a function of time for a $3 \times 3$ lattice. The three panels refer to different values of the coupling parameter $J_y/\gamma$ (the other parameters are $J_x/\gamma = 0.9$ and $J_z/\gamma = 1$).
}\label{fig:trajectories}
\end{figure}

In the three panels of Fig. \ref{fig:trajectories}, we show the results for $M^x_\Psi(t)$ in a $3 \times 3$ lattice for $J_y=0.95 \gamma$, $J_y=1.25 \gamma$ and $J_y=1.8 \gamma$. When the steady state presents a paramagnetic phase ($J_y=0.95 \gamma$, Fig. \ref{fig:trajectories}-(a)), the curve for $M^x_\Psi(t)$ presents only small fluctuations around the zero value for the magnetization. The behavior of the quantum trajectory is strikingly different in the ferromagnetic phase ($J_y=1.2 \gamma$, Fig. \ref{fig:trajectories}-(b)). In this case, we can clearly distinguish  intervals of time where the curve for $M^x_\Psi(t)$ fluctuates around a positive value of the magnetization and others where it fluctuates around the opposite value. The duration of these time intervals is of the order $\Delta t \sim \lambda^{-1}$.  Finally, for large values of the coupling parameter $J_y$ ($J_y=1.8 ,\gamma$, Fig. \ref{fig:trajectories}-(c)), $M^x_\Psi(t)$  presents yet another different behavior. It is reminiscent of what observed in the paramagnetic phase (see Fig. \ref{fig:trajectories}-(a)), since it fluctuates around the zero value of the magnetization, but the amplitude of the fluctuations is much larger than in the regime $J_x \simeq J_y$. This peculiar behavior can be ascribed to the strongly mixed character of the steady state in this regime (see Refs. \cite{BiellaPRX16,RotaPRB17} for a calculation of the purity and the von-Neumann entropy). In this case, the stochastic processes described by the increments $dW_j$ in Eq. \eqref{Eq:StocSchHomodyne} would allow the quantum trajectory to explore a much larger number of quantum states with respect to the case at small anisotropy, where the trajectory fluctuates weakly around the single pure state dominating in the steady-state density matrix. As a consequence, the fluctuations of $M_\Psi^x(t)$ in the paramagnetic regime of large anisotropy are much stronger than in the regime at $J_x \simeq J_y$.

\begin{figure}[h!]
\begin{center}
\includegraphics[width = 0.9\textwidth]{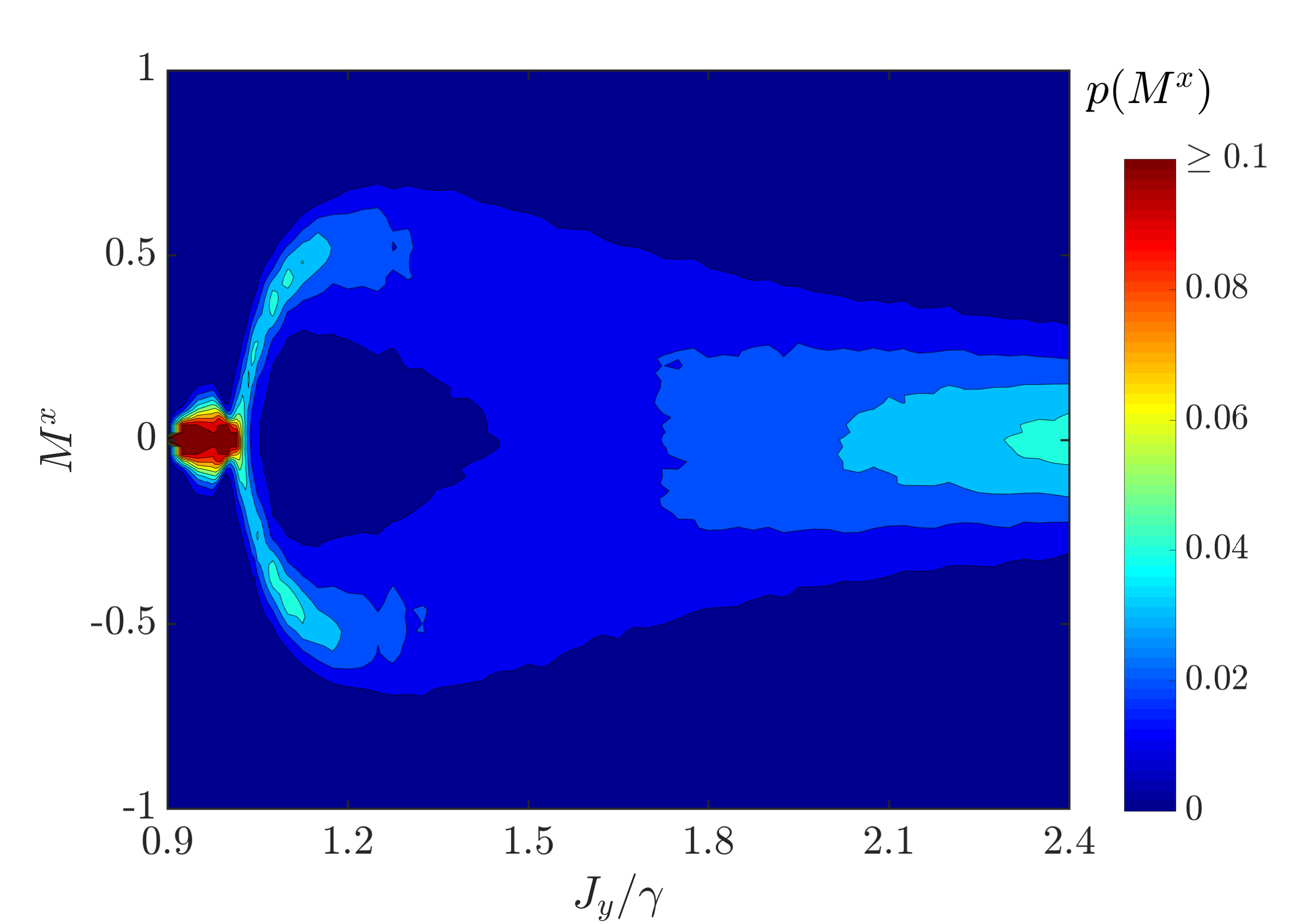}\\
\includegraphics[width = 0.9\textwidth]{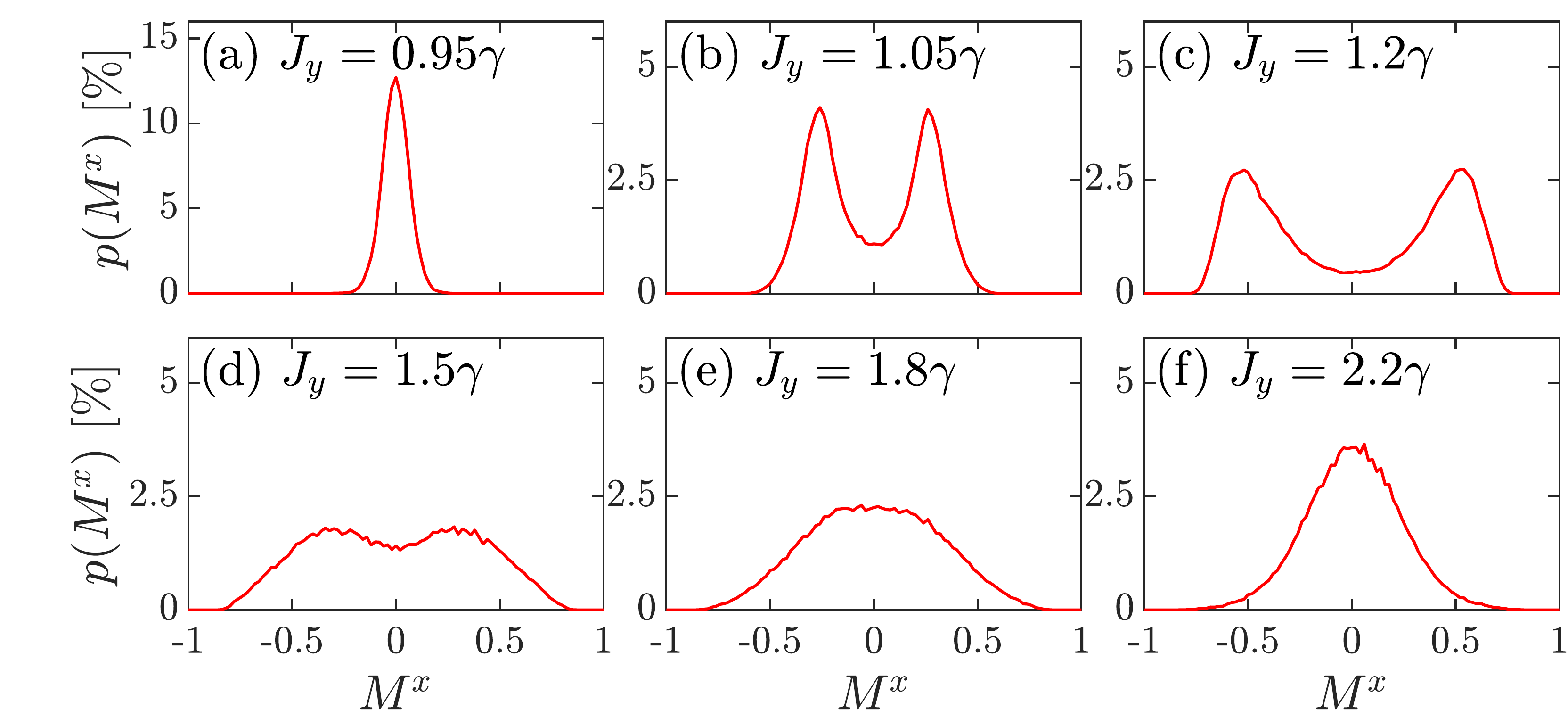}
\end{center}
\caption{Top panel: contour plot of the probability distribution $p(M^{x})$ of the site-averaged magnetization along $x$ versus the coupling parameter $J_y$ for a $3 \times 3$ lattice. Lower panels: probability distribution $p(M^x)$ for different values of $J_y$. 
For each value of $J_y$, the distributions are obtained collecting the results of $M^x$ from $N_T = 16$ trajectories with total time $t_T = 10^4 / \gamma$.
Same parameters as in Fig. \ref{fig:trajectories}.}\label{fig:distribution}
\end{figure}

To better understand the nature of those three regimes, we studied the probability distribution of $M^x_\Psi (t)$ over many trajectories, which we will call $p(M^x)$, defined as follows.
We consider a time $t_s$ where the density matrix of the system has reached the steady state, and statistically collect all the values of $M^x_\Psi (t)$ for $t > t_s$ over many trajectories. 
The results for $p(M^x)$ are presented in the top panel of Fig. \ref{fig:distribution}, as a function of the coupling $J_y$. We notice that for small $J_y$ the distribution is monomodal around zero. 
As $J_y$ increases, one reaches a point $J_c \simeq 1.05 \gamma$ where $p(M^x)$ starts to present two distinct peaks, which are symmetric around the value $M^x = 0$. If we continue to increase $J_y$, the two peaks broaden and they move apart, until they reach their maximum distance for $J_y \simeq 1.2 \gamma$. Above this value of $J_y$, the peaks continues to broaden and they start to approach one to the other, until they merge again into a single peak for $J_y\gtrsim 1.6 \gamma$. The broadening, the separation and the merging of the peaks in the probability distribution is even more evident in the panels in Fig. \ref{fig:distribution}-(a,f), where we plot the curves for $p(M_x)$ for some values of the coupling parameter $J_y$.

In order to perform a more quantitative analysis of the distribution $p(M^{x})$, we compute the bimodality coefficient $b$ \cite{Chissom70} that for an even distribution reads:
\begin{equation} b = \frac{(\int_{-1}^{1} dM_x M_x^2 p(M_x))^2}{\int_{-1}^{1} dM_x M_x^4 p(M_x)}.
\end{equation}
$b$ is an indicator of the bimodal character of the distribution, which in the present study is related to the ferromagnetic nature of the steady state. Indeed, when $p(M^{x})$ presents two narrow peaks, then the quantity $b$ approaches its maximum value $b_{max}=1$. Instead, unimodal distributions are characterized by smaller values of $b$ (for instance, a Gaussian distribution centered at $M_x = 0$ would have $b = 1/3$).

\begin{figure}[t]
\begin{center}
\includegraphics[width=0.6\textwidth]{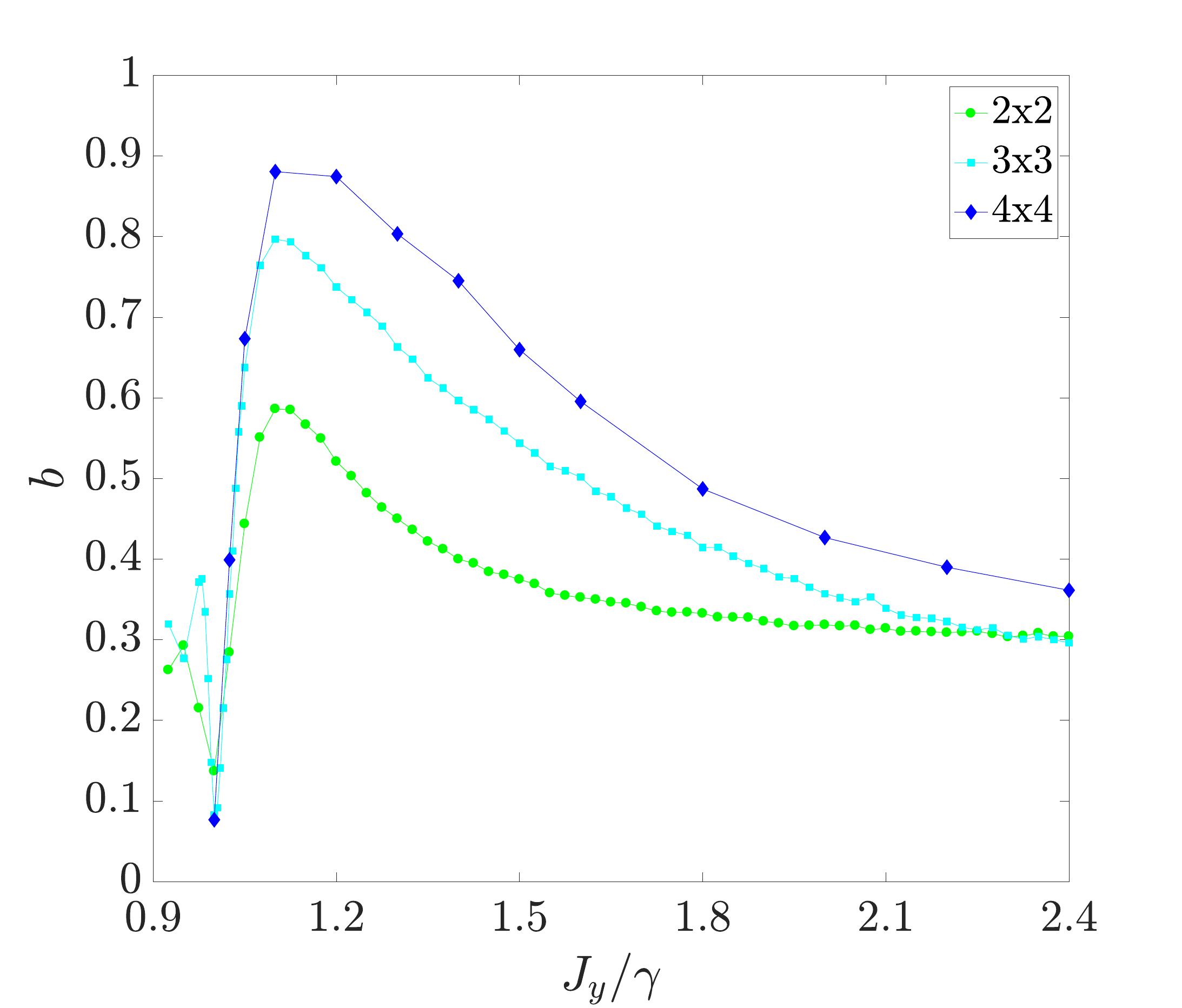}\\

\end{center}
\caption{Bimodality coefficient $b$ (defined in the text) as a function of the coupling parameter $J_y$, for different sizes of the 2D lattice. The full lines are a guide for the eye. Same parameters as in Fig. \ref{fig:trajectories}.}\label{fig:bimodality}
\end{figure}

In Fig. \ref{fig:bimodality}, we plot the value of $b$ as a function of $J_y$, for different sizes of the 2D lattice. The emergence of the phase transition at $J_y/\gamma \simeq 1.05$ is signaled by a steep increasing of the ratio $b$, which is almost independent of the lattice size. Furthermore, the decreasing of $b$ for $J_y/\gamma > 1.2$ indicates the disappearance of the ferromagnetic order for large anisotropies. In this case, however, the drop of $b$ is not particularly sharp and tends to become smoother and smoother as the size of the lattice increases.

The study of the behavior of $b(J_y)$ is interesting to address the open question about the nature of the steady state of the dissipative XYZ model for large anisotropies. Several works in literature \cite{BiellaPRX16,OrusNatComm17,CasteelsarXiv17} have predicted a ferromagnetic to paramagnetic phase transition for $J_y/\gamma > 1.5$. However, the critical value of $J_y$ for this second transition depends strongly on the method used and on the size of the cluster considered in the calculation \cite{BiellaPRX16,OrusNatComm17,CasteelsarXiv17}. Moreover, the behavior of the magnetic susceptibility and of the von-Neumann entropy as a function of $J_y$ do not present any feature signaling the emergence of a critical point for $J_y > 1.2 \gamma$ \cite{RotaPRB17}. Our results in Fig. \ref{fig:bimodality}, showing a smooth decreasing of $b$ at large $J_y$, together with the absence of a slowing down for $J_y > 1.2 \gamma$ (see Fig. \ref{fig:LiouvGap}-(d)), suggest that the disappearance of the ferromagnetic order for large anisotropies might be due to a crossover and not to another second-order phase transition.

\section{Conclusions}\label{sec:conclusions}

In this paper we investigated numerically the dynamics of a dissipative
spin-$\frac{1}{2}$ lattice interacting through an XYZ-Heisenberg Hamiltonian.
This model is particularly relevant in the context of strongly correlated open quantum systems since it is known to support a second-order dissipative phase transition in two dimensions, associated with the breaking of the $\mathbb{Z}_2$ symmetry.

By performing stochastic quantum trajectories simulations on finite-size
systems, we determined the Liouvillian gap from the asymptotic decay rate of the dynamics towards the steady state.
When the system is driven across the critical point, we found that the relaxation exhibits a slowing down. For 1D systems, the Liouvillian gap remains finite as the length of the chain is increased up to the thermodynamical limit, thus indicating the absence of a phase transition. Instead, results for 2D lattices do not show a saturation of the Liouvillian gap, which is consistent with the emergence of critical slowing down.
By analyzing individual stochastic homodyne trajectories in 2D lattices, we characterized the emergence and disappearance of two metastable states with opposite magnetization. Our predictions might be tested in quantum simulators based on superconducting quantum circuits or Rydberg atoms. As a perspective, the effects of disorder on the dynamics of these systems is a very interesting aspect that needs to be investigated in the future, as it is still unclear whether it can be detrimental to the emergence of the critical behavior \cite{Maghrebi17}, or if it may induce some other intriguing collective phenomena, such as many-body localization \cite{Nandkishore14,Levi16,Fischer16,Znidaric17,XuarXiv17}.  

\section*{Acknowledgments}
We thank N. Bartolo, F. Storme and F. Vicentini for useful discussions.  
We acknowledge support from ERC (via Consolidator Grant CORPHO No. 616233).

$^*$ The first two authors (R.R. and F.M.) contributed equally to this work.

%

\end{document}